\documentstyle[epsf]{article}
\newcommand{\be} {\begin{equation}}
\newcommand{\ee} {\end{equation}}
\newcommand{\bdm} {\begin{displaymath}}
\newcommand{\edm} {\end{displaymath}}
\newcommand{\bc} {\begin{center}}
\newcommand{\ec} {\end{center}}
\newcommand{\beqa} {\begin{eqnarray}}
\newcommand{\eeqa} {\end{eqnarray}}
\newcommand{\nn} {\nonumber}
\newcommand{\ra} {\rightarrow}
\begin{document}
\bc
{\Large\bf Hunting the Vector Hybrid}
\ec
\bc
{\large A Donnachie}\footnote{Permanent address: Department of Physics and 
Astronomy, University of Manchester, Manchester M13 9PL, England}
\ec
\bc
{\it TH Division, CERN, 1211 Geneva 23, Switzerland}
\ec
\medskip
\bc{\large Yu S Kalashnikova}
\ec
\bc
{\it ITEP, 117259 Moscow, Russia}
\ec
\bigskip
{\bf Abstract} The current state of analysis of $e^+e^-$ annihilation 
below 2.0 GeV and of the vector component of $\tau$-decay is reviewed. 
The evidence for and against the presence of hybrid vectors is discussed.
It is concluded that the data strongly favour their inclusion, and the
consequences of this are outlined. 
\bigskip

\bigskip
\section{Introduction}

It has been recognised for some time that the data on vector meson decays
appears to be in conflict with the predictions of the $^3P_0$ model which
has become the standard for calculating meson decays. One solution has been
to suggest that the physical light vectors are mixed $q\bar q$ and hybrid
vectors, as the latter have appropriate decay characteristics. However this
has never been quantified, nor have alternative non-hybrid explanations
been actively sought. Here we explore the limits of the $^3P_0$ model and
apply a specific non-$^3P_0$ model to the vector decays in an attempt to 
avoid the introduction of hybrids. Constraints placed on these models by other
decays, in particular those of the pseudoscalars $\eta(1295)$ and $\pi(1300)$,
are sufficient to prevent them from providing a solution to the vector decay
problem. Given that the inclusion of hybrids is unavoidable, we consider the
advantages and disadvantages of the flux tube model and the constituent gluon 
model of hybrids in the context of the vector decays. The data prefer the
constituent gluon model, and we outline briefly the consequences elsewhere
of this choice.

The current information on light-quark vectors from $e^+e^-$ annihilation 
and $\tau$ decay is discussed in Section 2, and the decay problems 
identified. Present understanding of gluonic excitations in general and of
hybrid models in particular is summarised in Section 3. In addition to the 
standard $^3P_0$ approach, which models the string breaking, we suggest a 
specific hadronic ansatz for relevant light quarkonia decays in analogy to 
the decays of heavy quarkonia. These two approaces are evaluated in the 
context of 
pseudoscalar decays, specifically $\eta(1295)$ and $\pi(1300)$, in Section 4 
and limits put on the corresponding vector decays. These latter results are
confronted with the data in Section 5, where it is shown that the limits 
are too restrictive to resolve the problems identified in Section 2. This
leads naturally to a detailed consideration of the two available models
for hybrids, and the data appear to favour the constituent gluon model
over the flux tube model. Our conclusions and their consequences are 
summarised in Section 6.

\section{$e^+e^-$ Annihilation and $\tau$ Decay}

The existence of the isovector $\rho(1450)$ and $\rho(1700)$, and their
isoscalar counterparts $\omega(1420)$ and $\omega(1600)$ is now well 
established \cite{pdg}. 

The key experimental results in determining the existence of the two 
isovector states were $e^+e^- \ra \pi^+\pi^-$ \cite{r1} and $e^+e^- \ra 
\omega\pi$ \cite{r2}. These original data sets have subsequently been 
augmented by data on the corresponding charged channels in $\tau$ decay 
\cite{r3,r4}, to which they are related by CVC. These new data confirm the 
earlier conclusions. The data on $e^+e^- \ra  \pi^+\pi^-\pi^+\pi^-$
\cite{r5} and $e^+e^- \ra \pi^+\pi^-\pi^0\pi^0$ \cite{r5,r6} (excluding 
$\omega\pi$) and the corresponding charged channels in $\tau$ decay 
\cite{r4} are compatible with the two-resonance interpretation \cite{r7,r8}.
However the $4\pi$ data alone do {\em not} provide such good discrimination
despite $4\pi$ being the major decay channel. The reason for this is 
straightforward. In $\omega\pi$ and $\pi\pi$ there is strong interference
with the tail of the $\rho$, which is absent in the case of $4\pi$. It
was also found that the $e^+e^- \ra \eta\pi^+\pi^-$ cross section
is better fitted with two interfering resonances than with a single
state \cite{r9}, with parameters in fair agreement with those found
in the analysis of other channels. Independent evidence for two 
$J^P = 1^-$ states was provided in a high statistics study of the
$\eta\pi\pi$ system in $\pi^- p$ charge exchange \cite{r10}. Decisive 
evidence for both the $\rho(1450)$ and $\rho(1700)$ in their $2\pi$ and
$4\pi$ decays has come from the study of $\bar{p} p$ and $\bar{p} n$ 
annihilation \cite{r11}.

The data initially available for the study of the corresponding isoscalar 
states $\omega(1420)$ and $\omega(1600)$ were $e^+e^- \ra \pi^+\pi^-\pi^0$
(which is dominated by $\rho\pi$) and $e^+e^- \ra \omega\pi^+\pi^-$ 
\cite{r12}. The latter cross section shows a clear peak which is apparently
dominated by the $\omega(1600)$. The former cross section shows 
little structure, but is appreciably larger than that calculated from the 
tails of the $\omega$ and $\phi$. This implies an additional contribution
and a best fit is obtained with two states \cite{r7}, although a fit with
only the $\omega(1600)$ cannot be excluded completely. Data on $\bar{p} N$
annihilation may help to clarify the situation, but analysis is still at a
preliminary stage \cite{r13}.

Although there is general consensus on the existence of the $\rho(1450)$,
$\rho(1700)$, $\omega(1420)$ and $\omega(1600)$ there is considerable
disparity on the parameters of these resonances. These show variation
from one reaction to another and, even within one particular process,
are dependent on the analysis techniques employed. Results 
from channels for which there is strong interference with the tail of 
the $\rho$ or of the $\omega$ and $\phi$ are sensitive to the choice of 
model used to estimate this contribution. For the $\rho(1450)$ 
the most extreme low mass comes from an analysis of the $\pi^+\pi^-$ 
spectrum in the reaction $K^- p \rightarrow \pi^+\pi^-\Lambda$ \cite{r14},
which gives $1266 \pm 14$ MeV. However such a low mass is not supported 
by any other analysis and does require confirmation. Most of the results
of the analyses of $e^+e^-$ annihilation, $\tau$ decay and $\bar{p} N$   
annihilation are clustered round the preferred PDG values \cite{pdg}, 
which are the ones we use here. These are given in Table 1.

\bc
\begin{tabular}{|c|c|c|c|c|}
\hline
Resonance & $\rho(1450)$ & $\rho(1700)$ & $\omega(1420)$ & 
$\omega(1600)$ \\
\hline
Mass(MeV) & $1465 \pm 25$ & $1700 \pm 20$ & $1419 \pm 31$ & 
$1649 \pm 24$ \\
\hline
Width(MeV) & $310 \pm 60$ & $240 \pm 60$ & $174 \pm 59$ & 
$220 \pm 35$ \\
\hline
\end{tabular}
\ec
\bc
Table 1: Experimental masses and widths of the higher vector mesons.
\ec

A natural explanation of these states is that they are the first radial, 
$2^3S_1$, and first orbital, $1^3D_1$, excitations of the $\rho$ and 
$\omega$ as the masses are close to those predicted by the quark model
\cite{gi}. This interpretation is given further credence by the 
observation of $\phi(1680)$ which has the appropriate mass to be
a candidate for the first radial excitation of the $\phi$. 

Despite the reasonable agreement of the observed masses with the quark
model predictions, the ratio of the $e^+e^-$ width of the $\rho (1700)$
to that of the $\rho (1450)$ is surprisingly large. In the non-relativistic
limit the $e^+e^-$ width of the $1^3D_1$ state vanishes, and although some
non-zero width will be created by relativistic corrections this is expected 
to be small. Additionally the data on the $4\pi$ channels in $e^+e^-$ 
annihilation and in $\tau$ decay do not appear at first sight to be 
compatible with those expected for the vector radial and orbital 
excitations of the $q\bar q$ system. This statement is of course model 
dependent as it assumes that we can predict the hadronic decays of the 
vector $q\bar q$ excitations. The $^3P_0$ model~\cite{3p0,ki,yopr,abs,bcps} 
does appear to allow this with some 
accuracy. A systematic study of known light $q\bar q$ decays shows that
a $^3P_0$-type amplitude dominates, and widths which are predicted to 
be large or small are found respectively to be so. More quantitatively, 
calculated widths agree with data to within $25 - 40 \%$.

The success of the $^3P_0$ model for well-known decays can be used to 
justify its application to predicting other decays, and in particular 
those of the radial and orbital excitations of the $\rho$ and $\omega$. 
In its simplest form the $^3P_0$ model contains only two parameters: an 
inverse length scale $\beta$ which controls the meson form factors, and 
the pair creation strength $\gamma$. These are not known precisely, but 
are reasonably well constrained with $\beta \sim 0.4$ GeV, $\gamma = 0.39$ 
GeV. Assuming that their masses are respectively $1.45$ and $1.70$ GeV, 
the $^3P_0$ partial widths for $\rho_{2S}$ and $\rho_{1D}$ are given in 
Table 2.
 
\bc
\begin{tabular}{|c|c|c|c|c|c|c|c|c|c|}  
\hline
Channel & $\pi\pi$ & $\pi\omega$ & $\rho\eta$ & $\pi h_1$ & $\pi a_1$ &
$\rho\rho$ & $\rho\sigma$ & other & total \\
\hline
$\rho_{2S}$ & 68 & 115 & 18 & 1 & 3 & 10 & 1 & 80 & 295 \\
\hline
$\rho_{1D}$ & 27 & 23 & 13 & 104 & 105 & 6 & 0 & 137 & 415 \\
\hline
\end{tabular}
\ec
\bc
Table 2: The $^3P_0$ partial widths for $\rho_{2S}$ and $\rho_{1D}$
\ec

In Table 2 ``other'' includes $K\bar K$, $K^*\bar K$ + c.c. 
and $6\pi$ channels, and the $\sigma$ is the broad S-wave $\pi\pi$ 
enhancement. Altogether 16 channels have been incorporated in the 
calculation~\cite{bcps}\cite{dktab}.

It is not necessary to go through a detailed analysis to show that these
$^3P_0$ model results exclude interpreting the $e^+e^-$ and $\tau$ decay 
data in terms of the $\rho_{2S}$ and $\rho_{1D}$ if the model is strictly 
applied. As already implied, the key is in the $4\pi$ decays. From 
Table 2 one can see that the $4\pi$ decays of the $\rho_{2S}$ are 
negligible, and so the $\rho_{2S}$ effectively makes no contribution to 
the $4\pi$ channel. In contrast the $4\pi$ decays of the $\rho_{1D}$ are 
large, and the two dominant ones, $h_1\pi$ and $a_1\pi$, are comparable. 
Now $h_1\pi$ contributes only to the $\pi^+\pi^-\pi^0\pi^0$ channel in 
$e^+e^-$ annihilation, but $a_1\pi$ contributes to both this and to 
$\pi^+\pi^-\pi^+\pi^-$. An immediate consequence is that we would expect
$\sigma(e^+e^- \ra \pi^+\pi^-\pi^0\pi^0) > \sigma(e^+e^- 
\ra \pi^+\pi^-\pi^+\pi^-)$, after subtraction of the 
$\omega\pi$ cross section from the total $\pi^+\pi^-\pi^0\pi^0$. This 
contradicts observation. Despite considerable uncertainty in the 
$\pi^+\pi^-\pi^0\pi^0$ cross section, enhanced by the
need to subtract the $\omega\pi$ cross section, it is undeniably 
appreciably smaller than the $\pi^+\pi^-\pi^+\pi^-$ cross section 
over most of the relevant energy range.

One explanation of this has been to suggest that the $q\bar q$ vector 
states are mixed with a hybrid vector~\cite{dk}\cite{cp} as this decays 
predominantly to $a_1\pi$ in flux tube models \cite{cp}, and to $a_1\pi$ 
and $\rho(\pi\pi)_S$ in constituent gluon models \cite{yulia}. 
Both the $\pi^+\pi^-\pi^+\pi^-$ and the $\pi^+\pi^-\pi^0\pi^0$ channels
are accessed by the $a_1\pi$ and $\rho(\pi\pi)_S$ decays so, in either case, 
$e^+e^-$ annihilation and the corresponding $\tau$ decays should in principle 
be explicable in terms of some suitable combination of $\rho$, $\rho_{2S}$, 
$\rho_{1D}$ and hybrid $\rho_H$, and with the implication that there
must be very little $\rho_{1D}$ to ensure the dominance of 
$\pi^+\pi^-\pi^+\pi^-$ over $\pi^+\pi^-\pi^0\pi^0$. The surprisingly large
ratio of the $e^+e^-$ widths is also a good indicator of mixing. 

However such evidence as we have from the isoscalar states indicates
that the picture might not be quite as simple as this. The $^3P_0$ widths 
for the $\omega_{2S}$ and $\omega_{1D}$ are given in Table 3 \cite{bcps}
assuming that their masses are respectively 1420 and 1650 MeV.

\bc
\begin{tabular}{|c|c|c|c|c|c|c|}
\hline
Channel & $\rho\pi$ & $\omega\eta$ & $b_1\pi$ & $\omega\sigma$ & 
Other & Total \\
\hline
$\omega_{2S}$ & 328 & 12 & 1 & 8 & 36 & 385 \\
\hline
$\omega_{1D}$ & 101 & 13 & 371 & 0 & 53 & 561 \\
\hline
\end{tabular}
\ec
\bc
Table 3: The $^3P_0$ widths for the $\omega_{2S}$ and $\omega_{1D}$
\ec

The large widths of the bare states predicted by the $^3P_0$ model are
well in excess of the quoted experimental total widths \cite{pdg}.
There must be strong mixing in the isoscalar channel as the $e^+e^-$ 
widths of the $\omega'_1$ and $\omega'_2$ are almost the same, and one 
would not expect either the $\omega_{1D}$ or the $\omega_{H}$ to have an 
electromagnetic coupling comparable to that of the $\omega_{2S}$. 
In the flux tube model the width of the $\omega_{H}$ is predicted to be 
small, $\sim 20$ MeV \cite{cp}, and is essentially all to $\rho\pi$. The 
$\omega_H$ width can be appreciably larger in constituent gluon models 
\cite{yulia} but again the $\rho\pi$ decay dominates although
some $\omega(\pi\pi)_S$ decay is allowed \cite{yulia}. Thus omitting
the $\omega_{1D}$, in analogy with the isovector case, would seem 
difficult to reconcile with the integrated cross section for
$e^+e^- \rightarrow \omega\pi\pi$ which, up to 1.8 GeV, is about
$60\%$ of the integrated $e^+e^- \ra \rho\pi$ cross section and
could be taken to imply some significant $\omega_{1D}$ component. 

The arguments relating to the $\rho(\pi\pi)_S$ and $\omega(\pi\pi)_S$
decays of $\rho_{2S}$ and $\omega_{2S}$ 
presuppose that there is no mechanism which can generate these
in any significant way. A possible approach is to invoke an inherent 
uncertainty in the $^3P_0$ model when applied to the decays of radial 
excitations to the ground state plus an $S-$wave $\pi\pi$ pair. In the 
$^3P_0$ model the decays of the $\rho_{2S}$ to $\rho (\pi\pi)_S$ and of 
the $\omega_{2S}$ to $\omega(\pi\pi)_S$ are strongly suppressed by a 
cancellation between two terms, one of which is strongly dependent on 
the model parameters. If these decays could be sufficiently enhanced 
within the structure of the model then the $4\pi$ problem in the isovector 
sector and the comparatively large $\omega\pi\pi$ in the isoscalar sector 
could possibly be resolved. Note that the $\rho(\pi\pi)_S$ and 
$\omega(\pi\pi)_S$ decays of the $\rho_{1D}$ and $\omega_{1D}$ respectively 
are strictly forbidden in the $^3P_0$ model. 

Further, many radial excitations are known to decay preferentially to the
ground state, or a lower radial excitation, plus $(\pi\pi)_S$.The most 
obvious ones occur in higher quarkonia. The branching fractions of
these decays are: $\psi(2S) \rightarrow \psi(1S)$, $50.8 \pm 3.7 \%$; 
$\Upsilon(2S) \rightarrow \Upsilon(1S)$, $27.3 \pm 1.4 \%$; $\Upsilon(3S)
\rightarrow \Upsilon(1S)$, $6.5 \pm 0.4 \%$, $\Upsilon(3S) \rightarrow
\Upsilon(2S)$, $4.8 \pm 0.7 \%$. These decays cannot proceed
via string breaking and are all specifically non-$^3P_0$
decays. A similar phenomenon is seen in light quarkonia: for example 
$\eta'(1295) \ra \eta(\pi\pi)_S$ and $\pi(1300) \ra \pi(\pi\pi)_S$, 
assuming for the moment that both $\eta'(1295)$ and $\pi(1300)$ are 
radial excitations. Both of these latter decays are essentially zero
in the $^3P_0$ model with standard parameters. Whether a similar 
mechanism is operating here as for heavy quarkonia, or whether
these decays arise from the sensitivity of the $^3P_0$ model for
radial decays involving $(\pi\pi)_S$, is undetermined. 

Before applying these various ideas to the vector meson decays we consider
the current status of hybrid mesons and of radial decays to the corresponding
ground state plus $(\pi\pi)_S$ in light quarkonia.  

\section{Hybrid Mesons}

Evidence for the excitation of gluonic degrees of freedom has
emerged in several processes. There are two independent indications
of an isovector $J^{PC} = 1^{-+}$ exotic resonance $\hat{\rho}(1600)$
in $\pi^- N \rightarrow \pi^+\pi^-\pi^- N$, specifically in the
$\rho^0\pi^-$ channel. The E852 collaboration~\cite{e852a} quote a
mass of $1593 \pm 8$ MeV and width of $168 \pm 20$ MeV, which are
consistent with the preliminary claim of the VES collaboration
\cite{vesa} of a resonance at $1620 \pm 20$ MeV with a width of 
$240 \pm 50$ MeV. There is also evidence for this state in the
$\eta'\pi$ channel~\cite{vesa}\cite{e852b}. It has been argued that 
the $\rho\pi$, $\eta'\pi$ and $\eta\pi$ couplings of this state
support the hypothesis that it is indeed a hybrid meson, although other
interpretations cannot be eliminated entirely~\cite{prp}. A peak in 
the $\eta\pi$ mass spectrum at 1.4 GeV with $J^{PC} = 1^{-+}$, in
the reaction $\pi^- N \ra \eta\pi^- N$,  has 
also been interpreted as a resonance~\cite{e852c}. Additional evidence 
for the same state in the same mode is provided by the Crystal Barrel 
collaboration \cite{cba}, in an analysis of $p\bar p \ra \eta\pi^+\pi^-$. 
In this case the signal is deduced from a phase variation in the 
$J^{PC} = 1^{-+}$ amplitude seen as interference in the Dalitz plot.
There is evidence from the VES collaboration \cite{vesb} for two isovector 
$0^{-+}$ states in the mass region 1.4 to 1.9 GeV. One is the well-established
$\pi(1800)$ \cite{vesc} with a mass of $1790 \pm 6 \pm 12$ MeV and width of
$225 \pm 9 \pm 15$ MeV, and one a new state, the $\pi(1600)$, with a mass
of $1580 \pm 43 \pm 75$ MeV and width of $450 \pm 60 \pm 100$ MeV. The quark
model predicts only one state in this mass region. Thus there is evidence
for degrees of freedom beyond $q\bar q$, and the unusual decay pattern 
of the $\pi(1800)$ encourages the belief that it has a strong hybrid
component \cite{bcps,cp}.

The interpretation of the peak in the $\eta\pi$ mass spectrum at 1.4 GeV 
as a resonance \cite{e852c} has been challenged \cite{dp}. It was shown 
that the E852 $\eta\pi$ peak and phase can be obtained without the need
to invoke the presence of an exotic resonance. The two key ingredients
are the presence of a strongly coupled threshold in this mass region
(taken to be $b_1\pi$) with rescattering to produce the $\eta\pi$
signal. A Deck-type background interfering with a hybrid resonance of 
higher mass, for which the $\hat\rho$ at 1.6 GeV is an obvious candidate,
was considered as the production mechanism. The Deck mechanism also provides 
the predominant natural parity exchange for the 1.4 GeV peak which is 
observed experimentally, in contrast to the 1.6 GeV state
which has a significant contribution from unnatural parity exchange.   
Of course the Deck mechanism is not applicable to the $p\bar p$ annihilation
experiment \cite{cba}, but the strongly-coupled threshold with rescattering 
can generate sufficient phase variation without requiring a resonance
at 1.4 GeV.

If the above is the correct interpretation of the $\eta\pi$ data then 
the $\hat{\rho}(1600)$ is the lowest mass exotic hybrid. An alternative 
viewpoint is to accept that the 1.4 GeV $\eta\pi$ signal really is an 
exotic resonance \cite{yulia} and to explore the consequences. 
One of these, which will be relevant for our subsequent discussion, is that 
the $J^{PC} = 0^{-+}$ $\pi(1300)$ is predominantly a hybrid meson, although
there may be some admixture of the $q\bar q$ $2^1S_0$ state which has the
same quantum numbers. One argument usually given in favour of the hybrid 
interpretation is the large $\pi(\pi\pi)_S$ branching fraction, much larger 
than $\pi\rho$. This is in complete disagreement with the predictions for 
the $q\bar q$ $2S$ level in the $^3P_0$ model for which the $\pi(\pi\pi)_S$ 
mode is strongly suppressed. However if the $\pi(1300)$ is a $q\bar q$
radial excitation then, as noted in Section 2, the decay is very sensitive
to the $^3P_0$ parameters. The decay may also be generated by a non-$^3P_0$
mechanism. Thus the $\pi(\pi\pi)_S$ decay may be possible without invoking
a hybrid. This possibility will be explored fully in Section 4.

Ideally we would have mass predictions for hybrids comparable to those
for the $q\bar q$ states. Unfortunately the absolute mass scale for 
light-quark hybrids is not precisely determined, with predictions for the
lightest hybrids lie between 1.3 and 1.9 GeV. Bag models \cite{bcv,cs} 
tend towards the lower end of this range, but it is not clear just how
reliable their results are. Parameters are tuned to fit the $q\bar q$ 
spectrum and it is questionable whether the same parameters should be 
used for the $q\bar{q}g$ states. Attempts to accomodate this lead to 
considerable variation in the predictions, giving a mass value for the 
lightest hybrid in the range 1.4 to 1.7 GeV. However it is perfectly 
possible to accomodate a $0^{-+}$ at $\sim 1.3$ GeV and a $1^{-+}$ at
$\sim 1.4$ GeV with the lightest $1^{--}$ hybrid at $\sim 1.65$ GeV.
Flux tube models \cite{ikp} predict hybrid masses to be considerably
higher than these, at about 1.9 GeV. The constituent gluon model
\cite{YuSK} gives the light-quark hybrid mass at 1.7 - 1.8 GeV.
In principle QCD sum rules could resolve the issue
of mass scale as they are a powerful tool for the understanding of 
hadron properties in terms of the vacuum condensates of QCD. However
even here there is a major divergence of view, either giving an upper
limit of 1.5 GeV on the mass of the $1^{-+}$ hybrid \cite{bdy}, with a 
preference  for a somewhat lower value, or putting it somewhere in the 
range 1.6 to 2.1 GeV \cite{lat}, with a preference for the upper end.
The principal difference between the two calculations is the application
of a low-energy theorem in \cite{bdy} which in turn gives an important 
r\^ole to the $g^3\langle G^3 \rangle$ term  with the effect of
lowering the mass. This is acknowledged in \cite{lat}.

However it does seem to be generally agreed that the mass ordering is 
$0^{-+} < 1^{-+} < 1^{--} < 2^{-+}$. This is certainly the case for bag 
models \cite{bcv,cs}, and also appears to hold in the heavy-quark sector
\cite{page98}. The same mass ordering emerges if one assumes that
the splittings are due to the spin-spin contact interaction \cite{yulia}.
The heavy-quark expansion of QCD in Coulomb gauge
\cite{ss} demonstrates that spin-orbit splitting of low-lying
hybrids with $J^{PC} = J^{-+}$ and $J^{+-}$ is such that $J = 1$
lies between $J = 0$ and $J = 2$, and the ordering is the same for
both sets. However either of $J = 0$ or $J = 2$ can be the lowest-lying. 
In lattice QCD calculations of heavy-quark hybrid states it is found that 
$0^{+-} < 2^{+-}$ \cite{laycock} so that combining the two results gives
$0^{-+} < 1^{-+} < 2^{-+}$ (and $0^{+-} < 1^{+-} < 2^{+-}$). Whether it 
makes sense to extrapolate these heavy-quark results to the light quark 
sector is debatable, but nonetheless the qualitative agreement with the  
bag and constituent model results is encouraging. 

The lack of precision in mass estimates is matched by uncertainty on
decay modes. Again the two standard approaches are the constituent
gluon model \cite{yulia,con} and the flux tube model \cite{cp,ikp}.
In the former the hybrids are considered specifically as having
three components: quark, antiquark and gluon. Decays proceed by 
dissociation of the constituent gluon \cite{con}. In the latter it
is assumed that the hybrids are quark-antiquark states moving on
an adiabatic surface generated by an excited flux tube of gluons,
with the standard $q\bar q$ mesons corresponding to the unexcited
flux tube. Decays of hybrids and $q\bar q$ mesons then proceed by the
same phenomenological pair-creation mechanism, for example the
$^3P_0$ model, coupled with a flux tube overlap \cite{ikp}. While
there are many common features in the decay modes predicted by these
approaches there are some substantial differences which become rather
crucial in interpreting data, and which are caused partly by the
different level of flexibility allowed within the models. In the constituent
model the decay strength is proportional to the strong coupling
constant $\alpha_s(q^2)$ given at some characteristic scale $q^2$. As the
present level of modelling does not permit definition of this scale, the
decay strength was treated as a model parameter in \cite{yulia,con}.
The analysis in \cite{yulia} was based on the assumption of $\hat{\rho}(1400)$
being a hybrid and some upper limits of its $\rho\pi$ mode, which were used
as an input to define the decay strength. 
In the original version \cite{ikp} of the flux tube model the decay strength
was defined from the data on $q\bar q$ decays. In principle, in the flux tube 
model the quark pair creation vertex is uncorrelated
with the gluonic modes of the hybrid. This permits the inclusion of
different decay vertices within the same overall structure. One has
recently been proposed \cite{pss}, motivated by the heavy-quark limit 
of the QCD Hamiltonian, and its predictions compared with those of
the standard $^3P_0$ vertex \cite{ss}. Once again there are many
similarities but some major differences which should be amenable
to experimental test although present data cannot distinguish. 

The hybrid decays of particular interest to us here are those of the
isovector $0^{-+}$ and $1^{-+}$, and both the isovector and isoscalar
$1^{--}$. There are some substantial differences between the flux tube 
model \cite{ikp,pss} and the constituent gluon model \cite{yulia,con} 
for these. We note that the $2S \rightarrow 1S + 1^3P_0$ decay has not been 
calculated in the flux tube model, although there is no {\it a priori} 
reason for it to be small apart from pure phase space considerations. If 
one identifies the bare $q\bar q$ $1^3P_0$ state with the physical $f_0(1300)$ 
then indeed there is not enough phase space to generate a sizeable width, 
and this seems to have been the approach adopted. However the standard quark 
model mass of the bare $q\bar q$ $1^3P_0$ state is about 1 GeV \cite{gi}, 
and the 
physical resonance position is shifted due to unitarity effects. This is 
the view we have adopted and it leads to an apparent difference between 
the flux-tube and the constituent gluon model for decays involving the 
$(\pi\pi)_S$ state.

{\bf $0^{-+}$:} In the flux tube model, the principal decay modes are 
$\rho\pi$ and, if the hybrid is sufficiently massive, $f_0(1370)\pi$. 
The $\rho\pi$ width in the constituent gluon model is comparable to that 
of the flux tube model, but it also has a very large $\pi(\pi\pi)_S$ width 
which dominates the decay.  

{\bf $1^{-+}$:} In this case the flux tube model and the constituent
gluon model are in reasonable accord. The principal decay modes are
$\rho\pi$ and $b_1\pi$, with the latter the larger of the two.

{\bf $1^{--}$:} The flux tube model predicts a rather narrow isovector state,
with $a_1\pi$ as the dominant mode. In contrast the constituent gluon model
predicts a much larger width, still with $a_1\pi$ dominant, but with
significant $\rho(\pi\pi)_S$ and $\omega\pi$ components. For the corresponding
isoscalar, the flux tube model predicts a very narrow state decaying almost
entirely to $\rho\pi$. Again the constituent gluon model predicts a much 
larger width, having $\rho\pi$ as the largest decay mode but also with
a significant $\omega(\pi\pi)$ fraction.

We would like to comment here on an important point concerning decays 
which include $(\pi\pi)_S$ in the final state. It has been argued that 
the complicated dynamics of the $(\pi\pi)_S$ final state is incompatible 
with the simple decay chain of $q\bar q$ in the $1^3P_0$ state going into 
$\pi\pi$. It has even been suggested that an effective Lagrangian approach 
may provide a better simulation of dynamics when the $(\pi\pi)_S$ system 
is involved \cite{frank}. Nevertheless, as the $1^3P_0$ $q\bar q$ couples 
strongly to $\pi\pi$, it should participate in the $\pi\pi$ dynamics even 
if there exists a non-$q\bar q$ mechanism which generates this dynamics. 
So there is no reasons to neglect the decay channels with $q\bar q$ in the 
$1^3P_0$ state, unless the corresponding amplitude is very small {\it per se}.

The interaction in the $(\pi\pi)_S$ channel is very strong and requires the 
unitarised coupled channel analysis, but the $\pi\pi$ phase shift can be 
described with the $1^3P_0$ $q\bar q$ as an intermediate state (see the 
detailed analysis of \cite{nils} and a simple model in Appendix A).
It is not surprising that a naive quark model, such as the $^3P_0$ one, fails 
to describe the low-mass part of the $\pi\pi$ $S$-wave phase shift, where
constituent quarks are not the proper degrees of freedom and chiral physics 
enters the game instead. It still remains an open question of how to 
incorporate the chiral symmetry constraints into the quark model unless it 
is done in a purely phenomenological way. In the simple model described in
Appendix A the $^3P_0$ amplitude is modified to interpolate smoothly between
the chiral perturbation theory regime with Adler zeroes and the confinement
regime with string-breaking modelled by the $^3P_0$ mechanism.

In summary there are two main lines which can be followed.

(i) Hybrids are comparatively light, the $\pi(1300)$ and $\hat\rho(1400)$
are hybrid states  (or, in the former case, predominantly hybrid) and the
mass of the hybrid $\rho_H$, $\omega_H$ $\sim$ 1.6 GeV. This would allow
strong mixing of the vector hybrids with the radial and orbital excitations
of the $\rho$ and $\omega$, but is not compatible with flux tube models.

(ii) Hybrids are comparatively heavy, the $\hat\rho(1600)$ is the lightest
$1^{-+}$ state and the $\pi(1600)$ presumably the corresponding $0^{-+}$ 
hybrid (or at least contains a significant hybrid component). This scenario
is compatible with flux tube models, but puts the $1^{--}$ vector hybrid mass
at $\sim$ 2.0 GeV, making strong mixing with the radial and orbital
excitations unlikely.

\section{Pseudoscalar Decays to Ground State plus {\bf $(\pi\pi)_S$}}

We have already commented that in the constituent gluon model of hybrids 
the decays of the $0^{-+}$ state differ from those predicted
by the flux tube model. It has been suggested \cite{yulia} that if the
$\hat\rho(1405)$ does exist then the $\pi(1300)$ should have a large
hybrid admixture and the $\pi(1300)$ decays would then allow a test of 
the two models. We explore both this hypothesis and the hypothsis that
the $\pi(1300)$ is a $q\bar q$ state, specifically the first radial 
$2^1S_0$ excitation of the $\pi$. For the latter assumption we can use
the decays of the $\eta(1295)$ as a control. To begin with we summarise
the current experimental situation with respect to the decays of both these
mesons. 

\subsection{$\eta(1275)$}

The most detailed study of the decay $\eta(1295) \ra \eta (\pi \pi)_S$ 
comes from the E852 charge-exchange reaction $\pi^- p \ra \eta\pi^-\pi^+ n$ 
\cite{manak}. The data are sufficiently precise to allow a separation of the 
$a_0 \pi$ and $\eta(\pi\pi)_S$ decays despite the similarity of these two 
channels. The $a_0\pi/\eta(\pi\pi)_S$ branching ratio is estimated to be 
$0.48 \pm 0.22$, although this may contain a large systematic error due to 
the difficulty of distinguishing unambiguously between the $a_0\pi$ and 
$\eta(\pi\pi)_S$ decays. This ratio disagrees with the GAMS result of $1.86
\pm 0.60$ \cite{gams}, although it also is possibly subject to similar
systematic errors. The total width of the $\eta(1295)$ is rather well defined:
$66 \pm 13$ MeV from E852, $53 \pm 6$ MeV from Fukui et al \cite{fukui}. 
Assuming that the two results for the $\Gamma(a_0\pi)/\Gamma(\eta(\pi\pi)_S)$ 
branching ratios give reasonable upper and lower limits, we can conclude that 
the partial width for $\eta(1295) \ra \eta(\pi\pi)_S)$ is in the approximate 
range 20 to 40 MeV.

\subsection{$\pi(1300)$}

Until recently there has been little information on the $\pi(\pi\pi)_S$
branching fraction of the $\pi(1300)$, and there is still considerable 
uncertainty in the total width, which can lie somewhere in the range 200 
to 600 MeV \cite{pdg}. The recent VES data \cite{vesb,vesc} show a clear
$\pi(1300)$ peak in $3\pi$, with a width of $\Gamma \sim 400$ to $500$
MeV in both $\pi(\pi\pi)_S$ and $\pi\rho$. The latter appears particularly 
strong and it has been suggested \cite{bcps} that as the size of the Deck
background in $\pi(\pi\pi)_S$ is uncertain it could provide the totality
of the $\pi(\pi\pi)_S$ signal. If this is correct then the dominant decay
would be $\rho\pi$. In contrast the E852 experiment \cite{e852d} claims
three decay modes of the of the $\pi(1300)$: $\pi\rho$, $\pi f_2(1270)$
and $\pi(\pi\pi)_S$. No comment is made on the mass or width of the 
$\pi(1300)$ other than it is broad. As for the VES experiment, the genuine
resonance signal could be confused by interference with the Deck background.

An uncertain Deck background is not a problem for $p\bar p$ annihilation
experiments. In their study of $f_0(1500)$ decays int $4\pi^0$ in $p\bar p
\ra 5\pi^0$ at rest, the Crystal Barrel experiment \cite{cbb} found a very
substantial improvement in their fit when the $\pi\pi(1300)$ decay of the
$f_0(1500)$ was included. As the final state in this case is all $\pi^0$
the decay of the $\pi(1300)$ {\em cannot} be to $\pi\rho$ and {\em must}
be to $\pi(\pi\pi)_S$. However because of the restriction to the purely
neutral channel nothing can be said about the $\pi(\pi\pi)_S$ branching
fraction. The parameters of the $\pi(1300)$, if left free in the fit, 
are determined to be $M = 1.114$ GeV, $\Gamma = 340$ MeV. Errors on these
are not quoted. The Obelix experiment \cite{oba}, in their analysis of
$p\bar p \ra 2\pi^+2\pi^-$ also find a significant improvement in their
fit if the $\pi(1300)$ is included in the decay chain. The $\pi(1300)$
parameters are found to be $M = 1.275 \pm 0.015$ GeV, $\Gamma = 218 \pm
100$ MeV. Both the $\pi(\pi\pi)_S$ and $\pi\rho$ modes were required
by the fit, and the ratio between them was found to be large:
$(\pi(1300) \ra \pi(\pi\pi)_S))/(\pi(1300)\ra \pi\rho) = 5.25 \pm 0.7$.
This is appreciably more than the result $\sim$ 2.12 obtained from
analysing much earlier data \cite{al}, and must be assumed to supersede it.
In conclusion, there seems to be little doubt that the decay $\pi(1300) 
\ra \pi(\pi\pi)_S$ not only exists, but is large and dominant. It is perhaps 
significant that the $p\bar p$ annihilation experiments find a smaller width 
than the $\pi p \ra (3\pi) p$ production experiments, which could be due to 
the effect of the Deck background on the latter. If this is the case, then
a total width of $\sim 300$ MeV would seem reasonable, with a partial
width for $\pi(\pi\pi)_S$ of more than 200 MeV.

\begin{center}
\begin{figure}
\begin{minipage}{100truemm}
\epsfxsize=125truemm
\epsffile{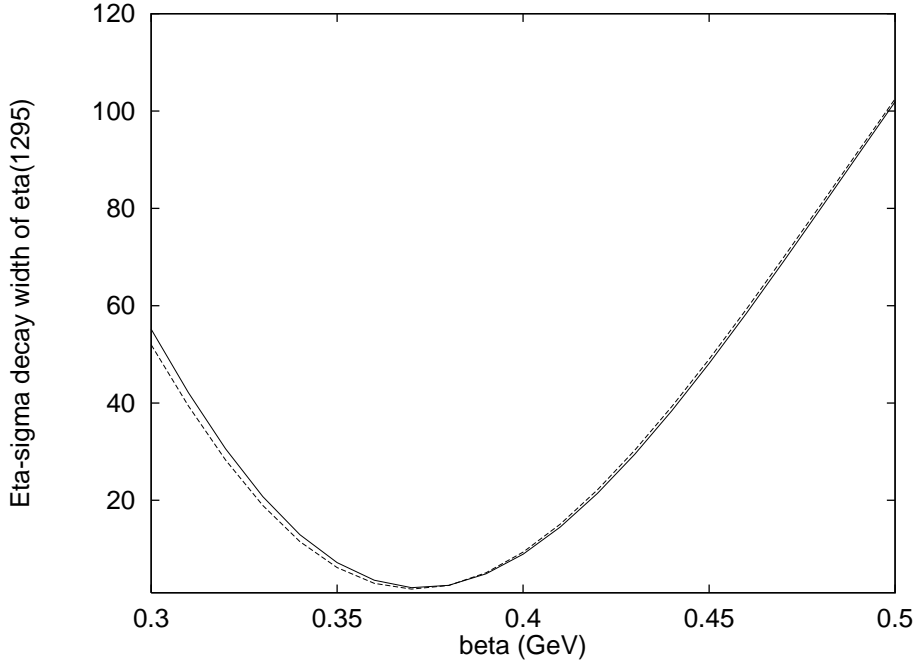}
\end{minipage}
\caption{Width of the decay $\eta'(1295) \rightarrow \eta (\pi\pi)_S$ as a 
function of $\beta$. The solid line is calculated using the $^3P_0$ $\pi\pi$
phase shift and the dashed line using the $^3P_0$ $\pi\pi$ phase shift 
modified to include the Adler zero.}
\end{figure}
\end{center}

\subsection{Pseudoscalar decays in the $^3P_0$ model}

For this part of the discussion we assume that the $\pi(1300)$ is a 
$q\bar q$ state, specifically the first radial $2^1S_0$ excitation of 
the $\pi$. For calculating decays in the $^3P_0$ model we also assume 
that the $(\pi\pi)_S$ is contained in that model and is the $1^3P_0$ state. 
A two-channel model which reproduces the experimental $S$-wave $\pi\pi$ 
amplitude is described in Appendix A. Both versions of the $(\pi\pi)_S$
amplitude described there were used in these calculations i.e. without and 
with the Adler zero included. The results are very much the same in both cases.

In the $^3P_0$ model, with standard wave-function parameters, the
decay of a radial excitation to the corresponding ground state plus 
$(\pi\pi)_S$ is small, a few MeV at most. This is true for $\rho_S$, 
$\omega_S$, $\eta(1295)$, $\pi(1300)$. It is caused by a node in the 
wave function, with the consequence that the decay is {\it very} 
sensitive to the parameter $\beta$. Of course there are limits within 
which $\beta$ can vary. A good guide to these limits is provided by 
\cite{abs}. Of the standard decays used to specify $\beta$ and $\gamma$, 
$a_1 \ra \rho\pi$ provides no constraint because of the very large 
experimental uncertainty on its width; $h_1 \ra \rho\pi$ provides only a 
weak constraint ($0.3\le \beta \le 0.5$ GeV), again because of experimental 
uncertainty coupled with a rather weak theoretical dependence on $\beta$; 
and $b_1 \ra \omega\pi$ also provides only a weak constraint, in this case 
because the theoretical width is almost independent of $\beta$ over quite 
a wide range. In contrast the theoretical widths of $f_2 \ra \pi\pi$, 
$a_2 \ra \rho\pi$ and $\rho \ra \pi\pi$ vary strongly with $\beta$. The 
two former favour a value of $\beta$ close to the mean of 0.4 GeV, but the 
latter prefers a much smaller value, $\sim 0.3$ GeV. In contrast the $D/S$ 
ratios for the decays $b_1 \ra \omega\pi$ and $a_1 \ra \rho\pi$, which 
are sensitive tests of the $^3P_0$ model, prefer a larger value, 
particularly the latter, which could go as high as $\beta \sim 0.5$ GeV. 
Taking everything into account, 0.3 GeV and 0.5 GeV do seem to provide extreme 
lower and upper limits on $\beta$, with 0.35 GeV and 0.45 GeV being more 
reasonable.

The variation of the $\eta(\pi\pi)_S$ width of the $\eta(1295)$ is given
in Fig.1 as a function of the value of $\beta$ for $(\pi\pi)_S$. Clearly 
the experimental limits on the $\eta(1295)$ width provide a strong constraint
on the allowed values of $\beta$, independently of other decays. 
In the $^3P_0$ model with standard values of the parameters $\beta$ and
$\gamma$ the width $\Gamma(\pi' \ra \pi\rho) \sim 200$ MeV and the width
$\Gamma(\pi' \ra \pi(\pi\pi)_S) \sim 0$ MeV \cite{bcps}. The $\pi\rho$
partial width does have quite a strong dependence on $\beta$, decreasing
from $\sim 300$ MeV at $\beta = 0.3$ GeV to $\sim 100$ MeV at $\beta \sim 0.5$
GeV \cite{bcps}. The variation of the $\pi(\pi\pi)_S$ width is much stronger,
due to the effect of the node in the wave function. This variation is
shown in Fig.2, where we have taken into account the effect of symmetrization
of like pions. The procedure for this is outlined in Appendix B. The 
constraints imposed by the $\eta(1295)$ decay width clearly restrict the 
maximum partial width for the $\pi(1300) \ra \pi(\pi\pi)_S$ to about 60 MeV.

So we conclude that the $^3P_0$ model can not be used to explain the
large $\pi(\pi\pi)_S$ width of the $\pi(1300)$. However before using 
this as an argument in favour of a hybrid interpretation we need to 
consider an alternative non-$^3P_0$ mechanism.

\begin{center}
\begin{figure}
\begin{minipage}{100truemm}
\epsfxsize=125truemm
\epsffile{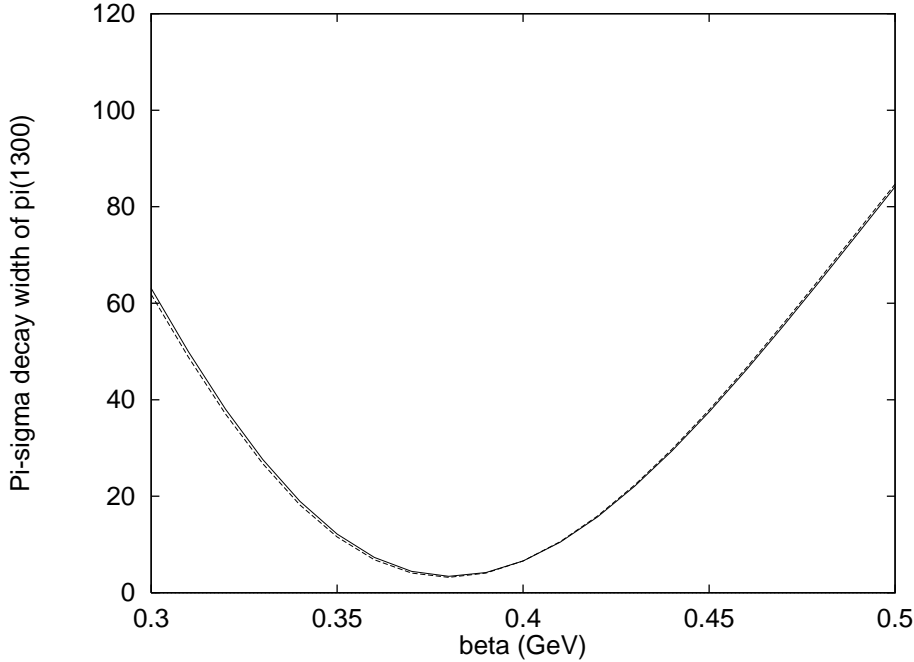}
\end{minipage}
\caption{Width of the decay $\pi(1300) \rightarrow \pi (\pi\pi)_S$ as a 
function of $\beta$. The curves are as in Figure 1.}
\end{figure}
\end{center}

\subsection{Non-$^3P_0$ decays of the pseudoscalars}

The decays of heavy-quarkonia radial excitations to the ground state, or
to a lower radial, plus $(\pi\pi)_S$ raise the possibility of analogous 
non-$^3P_0$ decays in the light-quark sector. The detailed mechanism is
presumably rather different from that of the models applied to the decays
of heavy quarkonia as they rely on a multipole expansion of the gluon field
\cite{hq}. The interactions of gluons with a wavelength $\lambda$ much 
larger than the radius $r_{Q\bar Q}$ of quarkonia are suppressed in the 
multipole expansion by powers of $r_{Q\bar Q}/\lambda$. So although it is
reasonable to consider only the leading operator for heavy quarkonia it is
not so for light quark states. 

An alternative mechanism which does not rely on a multipole expansion, 
``vacuum excitation'', has been suggested \cite{di}. This is
applicable in principle to both light and heavy quarkonia. The essential idea
is that a radial excitation decays to a lower radial excitation or to the
ground state by exciting a virtual state from the vacuum into reality. This
naturally has the quantum numbers of the vacuum, i.e. $I=0$, $J=0$ which
of course are precisely those of $(\pi\pi)_S$. For our present purposes
we simplify the calculation by considering relative phase-space as the 
spatial wave-functions of the $\eta(1295)$ and $\pi(1300)$ are identical, 
and so overlap integrals will be the same. We consider both unweighted and
weighted phase space, using for the latter the same $S$-wave $\pi\pi$ 
amplitude of Appendix A. It is irrelevant whether this is or is not a
genuine $q\bar q$ state. All that is required is an accurate representation
of the amplitude. We find that the ratio of $(\pi(1300) \ra \pi(\pi\pi)_S)/
(\eta(1295) \ra \eta(\pi\pi)_S)$ is 3.6 for unweighted phase space and
7.7 for weighted phase space. Thus given a width of 20 to 40 MeV for
the $\eta(1295)$ decay it is not difficult to generate the required large
width for the $\pi(1300)$ decay.

So we conclude that this particular non-$^3P_0$ mechanism can correlate
the $\eta(1295)$ and $\pi(1300)$ decays and provide a large $\pi(\pi\pi)_S$
width for the latter without the need to invoke a hybrid. Of course this
remains a hypothesis as we do not have a specific model with which to 
calculate these decays for light quarks.
 
\section{The Vector States}

We now consider the implications of the results of the previous sections
for the interpretation of the data from $e^+e^-$ annihilation and $\tau$
decay. We consider two extremes.

1. The vector hybrids are too heavy to permit significant mixing with
the $q\bar q$ states in the relevant kinematical region. The $\rho_S$ and

$\omega_S$ decays are some combination of $^3P_0$ and direct hadronic decay 
to $\rho(\pi\pi)_S$ and $\omega(\pi\pi)_S$ respectively. The $\rho_D$ and 
$\omega_D$ decays are purely $^3P_0$. Some mixing between the $2S$ and $1D$
states can be allowed.

2. The vector hybrids are sufficiently light to allow strong mixing
with the $q\bar q$ states. For completeness we consider the predictions of 
both the flux tube model and the constituent gluon model for the hybrid 
decays. However it must be remembered that the flux tube model prefers a 
higher mass for the vector hybrid.

\begin{center}
\begin{figure}
\begin{minipage}{100truemm}
\epsfxsize=125truemm
\epsffile{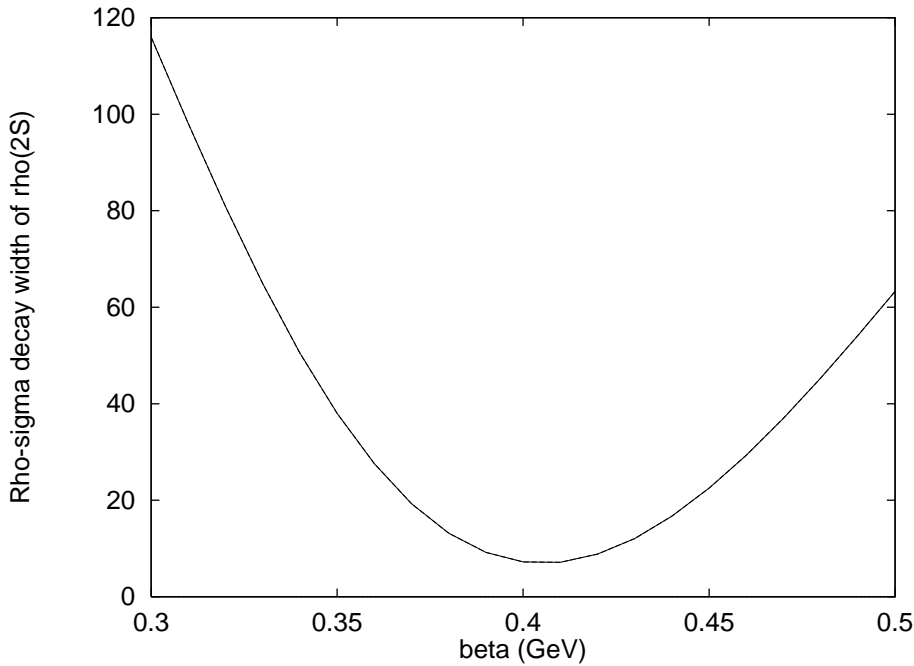}
\end{minipage}
\caption{Width of the decay $\rho (2S) \rightarrow \rho (\pi\pi)_S$ as a 
function of $\beta$. In contrast to Figures 1 and 2 only one curve is shown
as the results of using the different phase shifts are indistinguishable}
\end{figure}
\end{center}

\subsection{No Vector Hybrids}

The width of the decay $\rho_{2S} \ra \rho(\pi\pi)_S$ in the $^3P_0$ model 
is shown in Fig.3. Not surprisingly it is too small to account for the 
observed large $4\pi$ width of the $\rho(1450)$, even at the maximum
acceptable value of $\beta$. To extend the non-$^3P_0$ decay process 
suggested in Section 4.4 to the decay of the $\rho_{2S}$ requires the 
assumption that only $S$-waves are relevant. This is certainly in accord
with the decay of heavy quarkonia. The matrix element for the decay of
$\eta(1295)$ or $\pi(1300)$ is simply a constant, $f_S$ say. The general
form of the matrix element for the decay of $\rho_{2S}$ is
\bdm
A_{\mu\nu}=f\bigl\{-g_{\mu\nu}+{{k_{\mu}k_{\nu}}\over{k^2}}
+{{q_{\mu}q_{\nu}}\over{q^2}}
-{{(kq)k_{\mu}q_{\nu}}\over{k^2q^2}}\bigr\} 
+g\bigl\{{{(kq)k_{\mu}}\over{k^2}}-q_{\mu}\bigr\}\bigl\{k_{\nu}-
{{(kq)q_{\nu}}\over{q^2}}\bigr\}
\edm
where $k$ and $q$ are respectively the four-momenta of the $\rho_{2S}$ 
and the $\rho$. If only $S$-wave is present then 
\bdm
f = f_S \hskip 2truecm g = -f_{S}/m_0(m_1+E_1)
\edm
where $m_0$ and $m_1$ are the masses of the $\rho_{2S}$ and $\rho$, and
$E_1$ is the energy of the $\rho$ in the $\rho_{2S}$ rest frame. One has
then to calculate ${{1}\over{3}} A_{\mu\nu}^*A_{\mu\nu}$ which is simply 
$f_S^2$. So the only difference between the decays of the vector radial 
excitations and the pseudoscalar radial excitations is due to phase space. As 
the available phase space for the decay $\rho_{2S} \ra \rho(\pi\pi)_S$ is 
not very different from that for the decay $\eta(1295) \ra \eta(\pi\pi)_S$, 
the partial width will be comparable. 

Thus neither mechanism by itself can explain the $4\pi$ decays of the
$\rho(1450)$. Nor can one add them to increase the width, as the combined
strength is controlled by the $\eta(1295)$ decay. It should be recalled
that mixing of the $\rho_{2S}$ and the $\rho_{1D}$ is unlikely to resolve 
the difficulty as, even with an increase in the $\rho(\pi\pi)_S$ width
of the former within the limits allowed, the $\pi^+\pi^-\pi^0\pi^0$ cross
section will still exceed the $\pi^+\pi^-\pi^+\pi^-$ cross section. This
conclusion is not affected by adjusting the $^3P_0$ parameters for the
$\rho_{1D}$ decays, as the $a_1\pi$ and $h_1\pi$ widths move in unison
and the equality remains essentially unchanged. Thus we are forced to 
conclude that the $e^+e^- \ra 4\pi$ data can not be explained in terms of 
conventional $q\bar q$ dynamics.

The partial width for the decay $\omega_{2S} \ra \omega(\pi\pi)_S$ in either
of these models is comparable to that of the corresponding $\rho_{2S}$ decay.
This does not pose any particular problem for the isoscalar sector and does
not provide any further insight into the likely mechanisms as it is not
nearly sufficient in itself to provide the requisite integrated $\omega\pi\pi$ 
fraction. Thus we remain with the earlier conclusion that within the isoscalar 
$q\bar q$ structure there must be a significant $\omega_{1D}$ component. 

\subsection{Vector Hybrids}

It is now apparent that the inclusion of a isovector vector hybrid is 
essential to explain the $e^+e^- \ra 4\pi$ data, and consequently the
corresponding isoscalar vector hybrid must be included in any discussion
of the $e^+e^- \ra \rho\pi, \omega\pi\pi$ data. It is reasonable to assume
that the hidden-strange vector hybrid is sufficiently massive not to affect 
the discussion.

The real question is, what can be inferred about the nature of these hybrids?
The flux tube model with its dominant $a_1\pi$ decay would appear to resolve 
the problem in the isovector sector. To achieve strong mixing with the 
$\rho_{2S}$ and $\omega_{2S}$ requires a comparatively low mass. We have seen 
in Section 3 that the mass scale for the flux tube model is high, making 
strong mixing unlikely. However spin-dependent forces {\it may} lower the 
mass of the hybrid $\rho$ and $\omega$, which are spin $S = 0$ in contrast 
to the conventional $q\bar q$ components which are $S = 1$, sufficiently to 
allow strong mixing between hybrid and conventional quarkonia. There is
perhaps less of a problem with the constituent gluon model, but it is still
a concern. However accepting the low-mass hybrid scenario, one could 
then have
\bdm
\bigl\vert V \bigr\rangle = cos\phi \bigl\lbrace cos\theta
\bigl\vert 2^3S_1 \bigr\rangle + sin\theta \bigl\vert 1^3D_1
\bigr\rangle \bigr\rbrace + sin\phi \bigl\vert V_H \bigr\rangle
\edm
To explain the predominance of the $\pi^+\pi^-\pi^+\pi^-$ channel
in this model it is necessary to take $\theta \sim 0$, so that $\rho'_1$
is given by
\bdm
\bigl\vert \rho'_1 \bigr\rangle \sim cos\phi \bigl\vert \rho_S
\bigr\rangle + sin\phi \bigl\vert \rho_H \bigr\rangle
\edm
This simple scheme would make the $\pi^+\pi^-\pi^+\pi^-$ and 
$\pi^+\pi^-\pi^0\pi^0$ widths of the $\rho(1450)$ the same, with only 
the rather small $\rho(\pi\pi)_S$ decay of the $\rho_H$ contributing to 
the observed difference.
Further this suppression of the $\rho_{1D}$ contribution 
is not compatible with the isoscalar data. Without the $\omega_{1D}$ there is 
no source for the strong $\omega\pi\pi$ channel observed experimentally. 
The $\omega(\pi\pi)_S$ decay of $\omega_H$ is not sufficient to redress the
imbalance. The dynamics would be complex indeed if the 
$^3D_1$ state were absent, or nearly so, in the isovector channel but gave 
a significant contribution in the isoscalar channel. This is perhaps not 
entirely implausible as the isoscalar channel is complicated by the presence 
of the $\phi_{2S}$ right in the middle of the relevant mass range. Hadronic 
mixing with the nearby $\omega_{2S}$, $\omega_{1D}$ and $\omega_H$ will 
certainly occur at some level, modifying the isoscalar mixing pattern. 
However this mixing would have to be remarkably strong to produce the 
differences observed.

The opposite view, that the hybrid mass is high, also does not permit a
simple mixing scheme. Mixing essentially between the hybrid and the $^3D_1$ 
state is immediately ruled out by the $\rho(1450)$ decays, although
it would appear a sustainable option for the isoscalar hadronic decays.
Further the $e^+e^-$ widths of the $\rho(1700)$ and $\omega(1650)$ demand
mixing with the $\rho_{2S}$.  

The necessity to consider a complicated mixing scheme brings us back
to the problem of the hybrid mass scale. The simplest way to achieve strong
mixing between $\rho_{2S}$ and the hybrid is to have them nearly degenerate,
which means that the hybrid is very light, about 1.4 GeV. Setting the mass
scale in such a way opens up exciting possibilities for the spectroscopy of
states with nonexotic quantum numbers, along the lines discussed in 
\cite{yulia}. It has the additional advantage of being compatible with 
the sum rules result 
\cite{bdy}, which we consider as the most reliable of the sum rules analyses.
Nevertheless, the exotic $\rho$-$\pi$ signals at 1.6 GeV from BNL and VES 
appear sound, and an alternative explanation of the BNL signal at 1.4 GeV 
has been proposed. Moreover, it is very improbable to have two exotic 
hybrids so close to 
each other. Another argument against such a light hybrid comes
from the pseudoscalar sector: if there is no need to invoke a hybrid 
interpretation to explain the decay pattern of $\pi(1300)$, the new
pseudoscalars $\pi(1600)$ and $\pi(1800)$ from VES can be economically
considered as admixtures of $3S$ $q\bar q$ and hybrid states. It should 
be remembered that the hybrid interpretation of the $\pi(1800)$ \cite{bcps,cp}
emerges from its decay properties. 

An alternative scenario is to let the hybrid mix strongly with the
$1D$ state through their near-degeneracy, and then let the lower of the
two mixed states mix with the $2S$. This would mean that the $4\pi$
decays of the two observed physical states would be quasi-identical.
This also suggests that the third state is put rather high and, in addition, 
does not have much electromagnetic coupling. This scenario assumes
the vector hybrid at $\sim 1.7$ GeV, which makes the mass
problem easier to swallow and is compatible with the exotic and
pseudoscalar hybrid sector, including the splitting ordering. 

However this scenario is no longer the straightforward two-level 
mixing scheme initially proposed for the isovector channel. All three states
must be included, and in the isoscalar channel one cannot ignore the
possible additional complications arising from the $\phi_{2S}$. The
results of such mixing depend on the fine details of the positions of
the bare states and the mixing strengths. It is worth mentioning here that
one could expect rather strong mixing between quarkonia and hybrids in the
constituent model via constituent gluon emission/absorption. On the other
hand, there are no distinguishable gluons in the flux tube model, and,
consequently, no obvious mechanism to provide such mixing.
 
We have not attempted to construct a detailed mixing scheme here as it
requires consideration of all channels and a theoretically-constrained
fit to the data \cite{abc}. As a first step we present below a simple 
three-level mixing model which describes qualitatively the isovector data.

\subsection{A Simple Mixing Scheme}

We consider the mixing of the $2S$, the $1D$ and the hybrid $H_0$.
For the $3 \times 3$ mixing matrix use the standard PDG one without the
phase:
\be
\left(\matrix{c_{12}c_{13}&s_{12}c_{13}&s_{13}\cr
-s_{12}c_{23}-c_{12}s_{23}s_{13}&c_{12}c_{23}-s_{12}s_{23}s_{13}&
s_{23}c_{13}\cr
s_{12}s_{23}-c_{12}c_{23}s_{13}&-c_{12}s_{23}-s_{12}c_{23}s_{13}&
c_{23}c_{13}\cr}\right)
\ee
Then
\beqa
\vert \psi_1\rangle &=&c_{12}c_{13}\vert 2S \rangle +
s_{12}c_{13}\vert H_0 \rangle + s_{13}\vert 1D \rangle\nn\\
\vert H \rangle &=&-\left[s_{12}c_{23}+c_{12}s_{23}s_{13}\right]\vert 2S\rangle
+\left[c_{12}c_{23}-s_{12}s_{23}s_{13}\right]\vert H_0 \rangle + 
s_{23}c_{13}\vert 1D \rangle\nn\\
\vert \psi_2\rangle &=&\left[s_{12}s_{23}-c_{12}c_{23}s_{13}\right]\vert 
2S\rangle-\left[c_{12}s_{23}+s_{12}c_{23}s_{13}\right]\vert H_0 \rangle +
c_{23}c_{13}\vert 1D \rangle
\eeqa
We hypothesize that we want no direct mixing between the $\vert 2S \rangle$ 
and the $\vert 1D \rangle$ so set $s_{13} \sim 0$ and $c_{13} \sim 1$. Thus 
we have
\beqa
\vert \psi_1\rangle &=&c_{12}\vert 2S \rangle + 
s_{12}\vert H_0 \rangle\nn\\
\vert H \rangle &=&-s_{12}c_{23}\vert 2S \rangle + 
c_{12}c_{23}\vert H_0 \rangle + s_{23}\vert 1D \rangle\nn\\
\vert \psi_2\rangle &=& s_{12}s_{23}\vert 2S \rangle -
c_{12}s_{23}\vert H_0 \rangle + c_{23}\vert 1D \rangle
\eeqa
Obviously we identify $\vert \psi_1\rangle$ with the $\rho(1450)$. In the 
absence of a proper dynamical model of the mixing we are free to identify
$\rho(1700)$ either with $\vert H \rangle$ or with $\vert \psi_2\rangle$.
For definiteness we choose the former, but the subsequent discussion follows
analogously for the latter. 

It is reasonable to assume that the bare hybrid $\vert H_0 \rangle$ has no
direct electromagnetic coupling. We make no specific assumption on the
electromagnetic coupling of the bare $\vert 1D \rangle$ other than that it 
should be small compared to the $\vert 2S \rangle$. The $e^+e^-$ amplitudes 
for the physical states are
\beqa
\langle e^+e^- \vert \psi_1 \rangle &=& c_{12}\langle e^+e^- \vert
2S \rangle\nn\\
\langle e^+e^- \vert H \rangle &=& -s_{12}c_{23}\langle e^+e^- \vert 2S 
\rangle + s_{23}\langle e^+e^- \vert 1D \rangle\nn\\
\langle e^+e^- \vert \psi_2 \rangle &=& s_{12}s_{23}\langle e^+e^- \vert 2S 
\rangle +c_{23}\langle e^+e^- \vert 1D \rangle
\eeqa
We see immediately that provided $s_{12}c_{23} > 0$ (and $s_{23} > 0$
as the electromagnetic coupling of $\vert 1D \rangle$ has the oppposite sign 
to that of $\vert 2S \rangle$) we will get the correct relative signs of the 
electromagnetic couplings of the observed states \cite{r7}. In the limit of a 
vanishing 
electromagnetic width for the $\vert 1D \rangle$ then we require 
$s_{12}c_{23}$ to be $\sim {{1}\over{2}}$ to agree with the data analyses.

We know that the $\rho(1700)$ has a small $\omega\pi$ width, which puts
another constraint on the mixing. The $\omega\pi$ amplitudes are
\beqa
\langle \omega\pi \vert \psi_1 \rangle &=& c_{12}\langle 
\omega\pi \vert 2S \rangle\nn\\
\langle \omega\pi \vert H \rangle &=& -s_{12}c_{23}\langle \omega\pi 
\vert 2S \rangle + s_{23}\langle \omega\pi \vert 1D \rangle\nn\\
\langle \omega\pi \vert \psi_2 \rangle &=& s_{12}s_{23}\langle \omega\pi 
\vert 2S \rangle +c_{23}\langle \omega\pi \vert 1D \rangle
\eeqa
We have already established that $s_{12}c_{23} > 0$ and, in principle, that
$s_{23} > 0$. The latter is important as it allows here some cancellation 
between the two terms for the $\omega\pi$ decay of the $\vert H \rangle$. 
A very small width is thus not ruled out in principle, although in practice 
it may not be quite so simple as in the $^3P_0$ model the $\vert 2S \rangle$ 
width is very much larger than the $\vert 1D \rangle$ width.

Finally look at the $4\pi$ decays. We can ignore the $a_1\pi$ and $h_1\pi$
decays of the $\vert 2S \rangle$ but we can let it have some $\rho(\pi\pi)_S$
decay generated by other means. We should also let the bare hybrid have some
$\rho(\pi\pi)_S$ decay as well. The amplitudes are then:
\beqa
\langle a_1\pi \vert \psi_1 \rangle &=& s_{12}\langle 
a_1\pi \vert H_0 \rangle\nn\\
\langle a_1\pi \vert H \rangle &=& c_{12}c_{23}\langle a_1\pi \vert H_0 \rangle
+ s_{23}\langle a_1\pi \vert 1D \rangle\nn\\
\langle a_1\pi \vert \psi_2 \rangle &=& -c_{12}s_{23}\langle a_1\pi \vert 
H_0 \rangle +c_{23}\langle a_1\pi \vert 1D \rangle
\eeqa
and
\beqa
\langle h_1\pi \vert \psi_1 \rangle &=& 0\nn\\
\langle h_1\pi \vert H \rangle &=& s_{23}\langle h_1\pi \vert 1D \rangle\nn\\
\langle h_1\pi \vert \psi_2 \rangle &=& c_{23}\langle h_1\pi \vert 1D \rangle
\eeqa
and
\beqa
\langle \rho(\pi\pi)_S \vert \psi_1 \rangle &=& 
\langle \rho(\pi\pi)_S c_{12}\vert 2S \rangle 
+ s_{12}\langle \rho(\pi\pi)_S \vert H_0 \rangle\nn\\
\langle \rho(\pi\pi)_S  \vert H \rangle &=& 
-s_{12}c_{23}\langle \rho(\pi\pi)_S \vert 2S \rangle 
+ c_{12}c_{23}\langle \rho(\pi\pi)_S \vert H_0 \rangle\nn\\
\langle \rho(\pi\pi)_S \vert \psi_2 \rangle &=&
s_{12}s_{23}\langle \rho(\pi\pi)_S \vert 2S \rangle
-c_{12}s_{23}\langle \rho(\pi\pi)_S \vert H_0 \rangle
\eeqa
None of these are qualitatively inconsistent with observation. They all 
seem reasonable, particularly the lack of any $h_1\pi$ decay of
$\vert \psi_1 \rangle$. Note that $\vert \psi_2 \rangle$ has a non-zero
$e^+e^-$ width, and is presumably somewhere ``off-stage''. There is some
evidence for isovector states in the vicinity of 2.0 GeV which decay
strongly into $6\pi$ \cite{cd6}. One of these could be the missing member of 
the trio.
 
\section{Conclusions}

Our general conclusion is that the $e^+e^-$ annihilation and $\tau$ decay
data require the existence of a ``hidden'' vector hybrid in both the
isovector and isoscalar channels. The argument is based strongly on the
pattern of the observed decays to $\rho (\pi\pi)_S$ for the isovectors and
to $\omega (\pi\pi)_S$ for the isoscalars. The strong mixing evident from
the electromagnetic widths is also a key feature. The mixing is
non-trivial, involving the first radial and the first orbital $q\bar q$
excitations and the ground-state vector hybrid. 
Before coming to our general conclusion we explored the limits 
of the $^3P_0$ model and proposed a specific non-$^3P_0$ model in an 
unsuccessful attempt to explain these data without going beyond the 
$q\bar q$ sector.

More specifically we are inclined towards the constituent gluon model rather
than the flux tube model to describe the characteristics of the hybrids. This
is based on two aspects. Firstly the constituent gluon model can more readily 
encompass a  hybrid mass in the relevant range. Secondly we know that the
constituent gluon model yields strong decay modes involving $(\pi\pi)_S$,
while this is an unknown quantity for the flux tube model.

The suggestion that the ground state vector hybrid is comparatively light, 
i.e. about 1.6 GeV, has major consequences. Given the hybrid spin-parity 
sequence discussed in Section 3, namely $0^{-+} < 1^{-+} < 1^{--}$, it
reopens the question of a hybrid component in the $\pi(1300)$ \cite{yulia}
and emphasizes the urgency of clarifying the status of the $\hat\rho(1405)$.

{\bf Acknowledgements}

This research is supported in part by PPARC, by grant RFBR 96-15-96740 and by 
the EEC-TMR Program (contract NCT98-0169).

We are grateful to Frank Close for helpful discussions.
\vfill
\eject

{\bf Appendix A: The $(\pi\pi)_S$ Amplitude}

The simple $^3P_0$ model for the $\pi\pi$ $S$-wave amplitude assumes
that the scattering takes place via $0^{++}$ intermediate states. There
are two such states in this case, one is $\frac{1}{\sqrt{2}}(u\bar u +d\bar d)$
and the other is $s\bar s$.

The $\pi\pi$ phase shift $\delta$ and the inelasticity $\eta$ are given by
\bdm
tan(2\delta) = \frac{ab_1 - a_1b}{aa_1 + bb_1},
\edm
\bdm
\eta^2 = \frac{(aa_1+bb_1)^2 + (ab_1-a_1b)^2}{a^2+b^2},
\edm
where
\bdm
a = a(s) = 1 +\frac{1}{(s-s_1)(s-s_2)}
((Im\Delta_{12})^2 - Im\Delta_{11}Im\Delta_{22}), 
\edm
\bdm
b = b(s) = \frac{1}{s-s_1}Im\Delta_{11} + \frac{1}{s-s_2}Im\Delta_{22},
\edm
\bdm
a_1 = a_1(s) = 1 +\frac{1}{(s-s_1)(s-s_2)}
((Im\tilde{\Delta}_{12})^2 - Im\tilde{\Delta}_{11}Im\tilde{\Delta}_{22}), 
\edm
\bdm
b_1 = b_1(s) = \frac{1}{s-s_1}Im\tilde{\Delta}_{11} 
+ \frac{1}{s-s_2}Im\tilde{\Delta}_{22}.
\edm
In these formulae $\sqrt{s_1}$ and $\sqrt{s_2}$ are the masses of 
intermediate states, and
\bdm
Im\Delta_{ab} = \sum_{j}Im\Delta_{ab}(j), 
\edm
\bdm
Im\tilde{\Delta}_{ab} = Im\Delta_{ab} - 2Im\Delta_{ab}(\pi\pi),
\edm
where $a,b = 1,2$ label the intermediate states, and $j = \pi\pi, K\bar{K}$
are the coupled channels taken into account. 

The quantities $Im\Delta_{ab}(j)$ are calculated in the framework of $^3P_0$
model as
\bdm
Im\Delta_{ab}(j) = \frac{p_j}{8\pi\sqrt{s}}f_a(j)f_b(j)
\edm
with the $^3P_0$ form factors $f_a(j)$ defined as in \cite{ki}.

This model is an oversimplified version of a coupled channel method and
preserves unitarity but not analyticity; in contrast to the much
more sophisticated model \cite{nils} the physical resonance positions
$\sqrt{s_a}$ are taken to be constants and are not defined via dispersion
relations.

The $\pi\pi$ $S$-wave phase shift given by the $^3P_0$ model is shown as the 
solid line in Figure 4. To improve the agreement with the data close to the
$\pi\pi$ threshold the Adler zero is now introduced. We make it pragmatically
by substituting
\bdm
f_a(\pi\pi) \rightarrow \tilde{f}_a(\pi\pi) = f_a(\pi\pi) F(s)
\edm
with $F(s)$ being a smooth function of $s$. $F({{m_\pi^2}\over{2}}) = 0$,
$F(s) \rightarrow 1$ when $s$ increases. In such a way we have the Adler 
zero at $s_0 = {{m_\pi^2}\over{2}}$ in the $\pi\pi$ amplitude as required
by chiral symmetry. The $\pi\pi$ $S$-wave phase shift with this modification
is shown as the dashed line in Figure 4 for the simple choice
\bdm
F(s) = \theta(s_1 - s)\bigl\{
-\bigl({{\sqrt{s}-\sqrt{s_0}}\over{\sqrt{s_1}-\sqrt{s_0}}}\bigr)^2 + 
2\bigl({{\sqrt{s}-\sqrt{s_0}}\over{\sqrt{s_1}-\sqrt{s_0}}}\bigr)
\bigr\}
\edm
with $\sqrt{s_1} = 0.7$ GeV so that chiral dynamics and quark dynamics are
matched at 0.7 GeV.

The idea to insert Adler zeroes into the quark model form factors was suggested
in \cite{nils}. Our procedure differs from that one. In contrast to \cite{nils}
we believe that the soft pion physics governs only the lower end of the 
$\pi\pi$ mass spectrum and does not affect the whole mass range available,
leaving room for the strong interaction of quarks at higher energies.
Moreover it is our belief that with the more QCD-motivated model for hadronic
decays \cite{ss}, which takes into account the Goldstone nature of pions, the
quark model description is valid at lower $\pi\pi$ masses as well. This makes
the Adler zero constraints responsible for only a rather small energy range
just above the $\pi\pi$ threshold.

\begin{center}
\begin{figure}
\begin{minipage}{100truemm}
\epsfxsize=125truemm
\epsffile{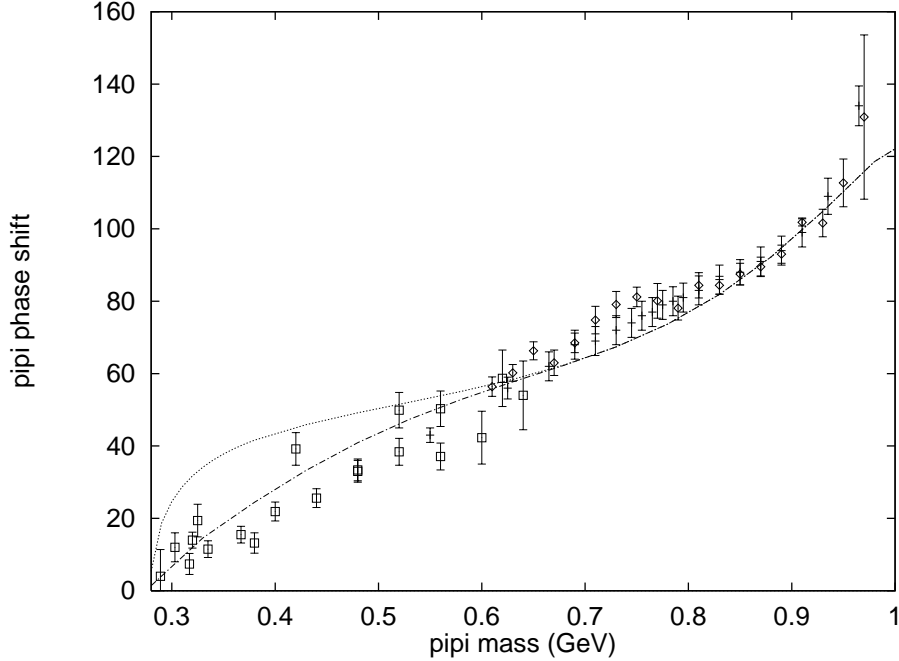}
\end{minipage}
\caption{The $S$-wave $\pi\pi$ phase shift. The solid curve is the result of 
the $^3P_0$ calculation and the dashed curve is the result of the low-energy
modification to include the Adler zero. The data are from \cite{data}}
\end{figure}
\end{center}

\vfill\eject

{ \bf Appendix B: The Piprime Decay Width}

The width of the decay of $\pi'^0$ into $\pi^+\pi^-\pi^0$ is given by
\bdm
d\Gamma = \frac{2}{3} \frac{1}{(2\pi)^3}\frac{1}{32m_{\pi'}^3}
|M_c|^2 ds_{12}ds_{13}
\edm
\bdm
|M_c|^2 = D
\edm
and the width into three neutral pions is
\bdm
d\Gamma = \frac{1}{3} \frac{1}{(2\pi)^3}\frac{1}{32m_{\pi'}^3}
|M_0|^2 ds_{12}ds_{13}
\edm
\bdm
|M_0|^2 = D + 2I
\edm
\bdm
D = f_{\pi'}^2(s_{12})f_{\sigma}^2(s_{12})P(s_{12})
\edm
\bdm
I = f_{\pi'}(s_{12})f_{\pi'}(s_{13})f_{\sigma}(s_{12})f_{\sigma}(s_{13})
P(s_{12})P(s_{13})
\edm
$$
((s_{12} - m^2_{\sigma})(s_{13} - m^2_{\sigma})
+ Im(s_{12})Im(s_{13}))
$$
\bdm
f_{\pi'}(s) = f_{\pi'\sigma\pi}(m_{\pi'}^2,s,m_{\pi}^2)
\edm
\bdm
f_{\sigma}(s) = f_{\sigma\pi\pi}(s,m_{\pi}^2,m_{\pi}^2)
\edm
\bdm
P(s) = [(s - m_{\sigma}^2)^2 + Im^2(s)]^{-1}
\edm
\bdm
Im(s) = \frac{1}{8\pi\sqrt{s}} p(s,m_{\pi}^2,m_{\pi}^2)f_{\sigma}^2(s)
\edm
The total widths of these decays are given by 
\bdm
\Gamma = \int^{s_{12max}}_{s_{12min}}\int^{s_{13max}}_{s_{13min}}
d\Gamma
\edm
\bdm
s_{13max} = \frac{m_{\pi'}^2 + 3 m_{\pi}^2 - s_{12}}{2}
+ 2\sqrt{\frac{(s_{12}-4m_{\pi}^2)}{4}\frac{(m_{\pi'}^2 - m_{\pi}^2 -s_{12})^2
- 4m_{\pi}^2s_{12}}{4s_{12}}}
\edm
\bdm
s_{13min} = \frac{m_{\pi'}^2 + 3 m_{\pi}^2 - s_{12}}{2}
- 2\sqrt{\frac{(s_{12}-4m_{\pi}^2)}{4}\frac{(m_{\pi'}^2 - m_{\pi}^2 -s_{12})^2
- 4m_{\pi}^2s_{12}}{4s_{12}}}
\edm
\bdm
s_{12max} = (m_{\pi'} - m_{\pi})^2
\edm
\bdm
s_{12min} = 4m_{\pi}^2
\edm

\vfill\eject

\vfill\eject

\end{document}